\documentclass[aps,twocolumn,amsmath,amssymb,preprintnumbers,floatfix,prl,longbibliography]{revtex4-2}

\usepackage{comment}
\usepackage[version=4]{mhchem}
\usepackage[utf8]{inputenc}
\usepackage{newtxtext}
\usepackage{microtype}
\usepackage{textcomp}
\usepackage{dsfont}
\usepackage{eucal}
\usepackage{siunitx}
\usepackage{soul}
\usepackage{tikz}

\usepackage{enumerate}
\usepackage{amsfonts}
\usepackage{color}
\usepackage{soul}
\usepackage{mathrsfs}

\usepackage{todonotes}
\presetkeys%
    {todonotes}%
    {inline}{}

\usepackage{graphicx}

\usepackage[colorlinks,allcolors=blue]{hyperref}
\usepackage[capitalize]{cleveref} 
\usepackage{cleveref}

\newcommand{\expect}[1]{\langle #1 \rangle}                     

\newcommand{\mr}[1]{\mathrm{#1}}

\definecolor{DarkBlue}{rgb}{0,0,0.80}
\definecolor{DarkRed}{rgb}{0.80,0,0}
\definecolor{DarkGreen}{rgb}{0,0.80,0}
\definecolor{Purple}{rgb}{0.55,0,0.55}

\newcommand{\up}{\uparrow}                                      
\newcommand{\dn}{\downarrow}                                    

\newcommand{\ie}{i.e.\ }

\let\epsilon\varepsilon

\begin{document}

\title{Real-space microscopic description of laser-pulse induced melting of superconductivity}
\author{Karl Bergson Hallberg}
\affiliation{Center for Quantum Spintronics, Department of Physics, Norwegian \\ University of Science and Technology, NO-7491 Trondheim, Norway}

\author{Guillermo Nava Antonio}
\affiliation{Cavendish Laboratory, University of Cambridge, CB3 0HE Cambridge, UK}

\author{Chiara Ciccarelli}
\affiliation{Cavendish Laboratory, University of Cambridge, CB3 0HE Cambridge, UK}
\affiliation{Center for Quantum Spintronics, Department of Physics, Norwegian \\ University of Science and Technology, NO-7491 Trondheim, Norway}\date{\today}

\author{Jacob Linder}
\affiliation{Center for Quantum Spintronics, Department of Physics, Norwegian \\ University of Science and Technology, NO-7491 Trondheim, Norway}\date{\today}

\begin{abstract}
	Quenching quantum order via laser pulses has proven a useful tool to access
	exotic physical effects in systems that are strongly perturbed out of equilibrium. 
    However, theoretical modelling of experimental measurements is typically done
	phenomenologically or by assuming translational invariance due to the complexity
	of the problem. Here, we solve a microscopic real-space model of the time dynamics
	of a superconductor following an intense laser-pulse. We are able to reproduce
	recent experimental findings displaying a critical slowing-down of the melting
	of the order parameter for laser fluences close to the condensation energy.
	Moreover, we leverage the real-space resolution of our model to predict how
	phase fluctuations and currents in the system behave both spatially and temporally.
	We discover an unusual current flow in the superconductor after the pulse has
	subsided, resembling backward waves that normally require special engineering
	in metamaterials or wave guides. Our results predict a rich behavior of the
	superconducting order parameter at a microscopic level which is manifested in
	current textures that can be probed using radiation detection.
\end{abstract}

\maketitle
\textit{Introduction}. Ultrafast optical pump-probe experiments have made it possible
to observe superconductivity melting and recovering on timescales on the order
of picoseconds
\cite{giannetti_aip_2016, kampfrath_natphot_2013, koch_natrevmet_2023, nemec_jchemphys_2005}.
These experiments allow one to visualize the melting and subsequent recovery of
superconductivity revealing complex dynamics and non-monotonic behavior with
regards to pump fluence \cite{beck_prl_2011, matsunaga_prl_2013, beck_prl_2013}.
The measurements probe a regime far out of equilibrium where a complex interplay of
the various degrees of freedom determines observable phenomena such as melting
and recovery times \cite{matsunaga_prl_2012, kabanov_prl_2005}.

Despite the experimental studies, a fully microscopic theoretical description
capable of connecting the microscopic pairing physics to the experimentally
accessible signals under strong time-dependent fields is lacking, in particular
including spatial resolution of the dynamic superconducting order parameter.
Phenomenological models such as Ginzburg-Landau give less intuition about the
microscopic degrees of freedom and are not formally applicable far out of equilibrium 
and other models often work under the assumption of translational invariance or
the electromagnetic field having a frequency less than that of the superconducting
gap which is insufficient to completely destroy superconductivity
\cite{papenkort_prb_2007, papenkort_jconfser_2009, zachmann_njp_2013, tanabe_prb_2018, hohenberg_revmodphys_1977, goldenfeld_lectures_1992}.

In this Letter, we develop a microscopic, spatially-resolved, real-time and
real-space description of the pump induced suppression and recovery of
superconductivity on a 2D lattice. We investigate the dynamics of a
superconductor subjected to a high-frequency laser pulse using self-consistent
time-dependent Bogoliubov-de Gennes (BdG) equations and the Heisenberg equations
within a tight-binding framework. This method allows for a spatially resolved
treatment of the order parameter and the inclusion of the electromagnetic field
via the Peierl's substitution modifying the hopping parameter, and thus opens
the possibility to study ultrafast dynamics in heterostructures. Our microscopic
model is able to reproduce the experimental observation \cite{guillermo_arxiv_2026} of a critical
slowing-down of the superconducting melting for laser fluences close to the condensation
energy. Moreover, from the real-space resolution of our simulations, we discover
that the temporal and spatial phase fluctuations in the system produce an unusual
current flow pattern in the superconductor after the duration of the laser pulse.
The current flow has opposite phase and group velocity and is thus a physical realization
of a backward wave \cite{meitzler_asa_65} that normally requires special engineering
of metamaterials \cite{hummelt_ieee_14} or wave guides \cite{ibanescu_prl_04}.
Our results thus uncover different and rich behavior of the superconducting order
parameter at a microscopic level, which can be experimentally probed using radiation
detection.
\textit{Theory.} The superconductor is modelled by the
following mean-field Hamiltonian defined on a $N_x \times N_y$ lattice
\begin{equation}
	\begin{split}
		 & \hat H = \hat H_h + \hat H_\Delta + \hat H_{\mathrm{ph}} +
		\hat H_{\mathrm{e-ph}} + \hat H_\mu + H_0
		\\ &= \sum_{ij \sigma} (c_{i\sigma}^\dag t_{ij}^{\sigma \sigma'}(\tau) c_{j\sigma'} +
		\text{H.c}) + \sum_i (\Delta_i c_{i\up}^\dag c_{i\dn}^\dag
		+ \Delta_i^* c_{i\dn}c_{i\up})
		\\ &+ \sum_i \omega_{\mathrm{ph}} b_i^\dag b_i
		+ \gamma \sum_{i\sigma}c_{i\sigma}^\dag (b_i^\dag + b_i)c_{i\sigma}
		+ \sum_i \mu_i n_i^f + H_0,
	\end{split}
\end{equation}
where $\tau$ is time, $\Delta_i \equiv U \expect{c_{i\dn}c_{i\up}}$ is the
superconducting order parameter, $\omega_{\mathrm{ph}}$ is the frequency of the
phonons, $\gamma$ the strength of the coupling between fermions and phonons,
$\mu_i$ is an onsite chemical potential,
$n_i^f \equiv \sum_\sigma c_{i\sigma}^\dag c_{i \sigma}$ and
$H_0 = \sum_i |\Delta_i|^2/ |U|$.
We employ Peierl's substitution which entails
modifying the hopping term to include a time-dependent phase, which for a laser
pulse with a spatial wavelength much larger than the relevant size
of the superconductor simplifies to
\begin{equation}
	t_{ij}(\tau) =
	\begin{cases}
		-t\mathbb I, \quad j = i + y \\
		-te^{\mathrm i aA_x(\tau)}\mathbb I, \quad j = i + x,
	\end{cases}
\end{equation}
where $a$ is the lattice constant and the vector potential is given by
\begin{equation}
	A(\tau) = \Theta(\tau)A_0
	e^{-\frac{(\tau - \tau_p)^2}{2 \sigma_p^2}}\sin \Omega_{pf} \tau \; x.
\end{equation}
The parameters used for the simulation are $t = a = \hbar = 1$, $N_x, N_y \in \{100, 120\}$,
$U = -1$, $\omega \in \{0, 0.3\}$, $\gamma \in \{0, 0.05\}$, $\mu \in \{0, 0.3\}$,
$A_0 \in [1, 3]$, $\sigma_p = 0.25$, $\tau_p = 5\sigma_p$, $\Omega_{pf} = 9.44$.
The last four parameters were chosen to correspond to experiment.

\textit{Time dynamics and initial conditions.} To investigate the effect of the
laser pulse on the superconducting order parameter we are interested in the time
evolution of correlators of the system such as $\expect{c_{i \dn} c_{i \up}}(\tau)$.
These correlations are calculated via the Heisenberg equation,
\begin{equation}
	i\frac{d}{d\tau}\hat O_H(\tau) = [\hat O_H(\tau), \hat H(\tau)],
\end{equation}
which lead to a large set of coupled non-linear differential equations. The Heisenberg
equations allows us to track the time dependency of all the relevant microscopic
correlations at a lattice resolution. Even though the Hamiltonian \textit{a priori}
only includes nearest neighbor hopping and on-site superconducting pairing, the
commutator of \textit{e.g.} $c_{i\dn} c_{i \up}$ and the hopping term yields time
dynamics dependent on arbitrary off-diagonal correlations $c_{i \dn} c_{j \up}, \; i \neq j$.
To produce a reasonable set of initial conditions, we assume that the system before
the laser pulse is in equilibrium such that we may perform a mean-field approximation
on the phonons. The derivation and the full sets of equations of motion are given
in the supplemental material \cite{supplemental}.

\textit{Absorbed energy and melting time.} A recent experiment \cite{guillermo_arxiv_2026}
showed that the time dynamics of the order parameter $\Delta$, and in particular
its melting, was strongly dependent on the absorbed energy due to the injected laser
fluence. A critical slowing down of the order parameter melting was observed when
the absorbed energy equaled an amount that was larger than the condensation energy
of the superconductor, but smaller than the energy required to completely melt the
superconducting order, $\Delta \to 0$. To compare the energy that is absorbed
by the superconductor as a result of the laser in our model, we calculate the
internal energy $\expect{\hat H}(\tau)$ with vector potential $A(\tau) = 0$
during the time evolution. $A$ is set to zero because we are interested in the
energy that is absorbed by the superconductor and not the energy associated with
the vector potential which will vanish when the pulse is over. In this way only
the energy that is absorbed by the system after the pulse subsides is accounted
for \cite{cui_prb_2019}. The internal energy is therefore given by 
\begin{equation}
	\begin{split}
		\expect{\hat H[A = 0]} = \sum_{ij} \mr{tr}(t^\mr s_{ij}\rho_{ij})
		+ \sum_i \left( \Delta_i \Gamma_{ii}^{\up\dn}
		+ \Delta_i^* F_{ii}^{\dn\up}\right) \\
		+ \sum_l \omega_{\mr{ph}} \expect{n^b_l}
		+ \sum_i \left (\gamma X_i + \mu_i \right) \expect{n^f_i} + H_0,
	\end{split}
\end{equation}
where the trace is over the spin degrees of freedom. 

We track the change in this value at time $\tau$ to its value at time $\tau = 0$,
\ie the change in, or absorbed, internal energy
\begin{equation}
	\mathcal E(\tau) \equiv \langle \hat H[A = 0] \rangle (\tau) - \langle \hat H[A = 0] \rangle (0).
\end{equation}
This value may in turn be compared to the difference in internal energy between the
superconducting state and normal state in equilibrium before the pulse is turned on,
\begin{equation}
	\mathcal E_{\mr{sc}} \equiv |\langle \hat H \rangle_{\mr{sc}}(0) - \langle \hat H \rangle_{\mr{n}}(0)|,
\end{equation}
which provides a sense of scale of the energy absorbed from the laser pulse.
After the pulse is turned on, the system is no longer in equilibrium, therefore
we do not have access to equilibrium values such as temperature, and we cannot
meaningfully calculate values such as the free energy. Therefore we find that
it is more meaningful to compare the change in internal energy to the difference
in internal energy between the superconducting state and the normal state before
the quench, and not the condensation energy given by the difference in free energy
between the superconducting and normal state. This is because the free energy
also includes a contribution from entropy $S$. However, at very low temperatures
$T$, the condensation energy and internal energy difference at equilibrium coincide
since $TS \to 0$. As for the melting time $\tau_m$, it is defined as the time it
takes for the site-averaged order parameter to reach within a percentage of its
minimum value, in this Letter we have used $1\%$. This is to reduce the influence
of numerical inaccuracies and small oscillations, especially in cases where the
order parameter changes very little from its original value.

\begin{figure}
	\centering
	\includegraphics[width=1\linewidth]{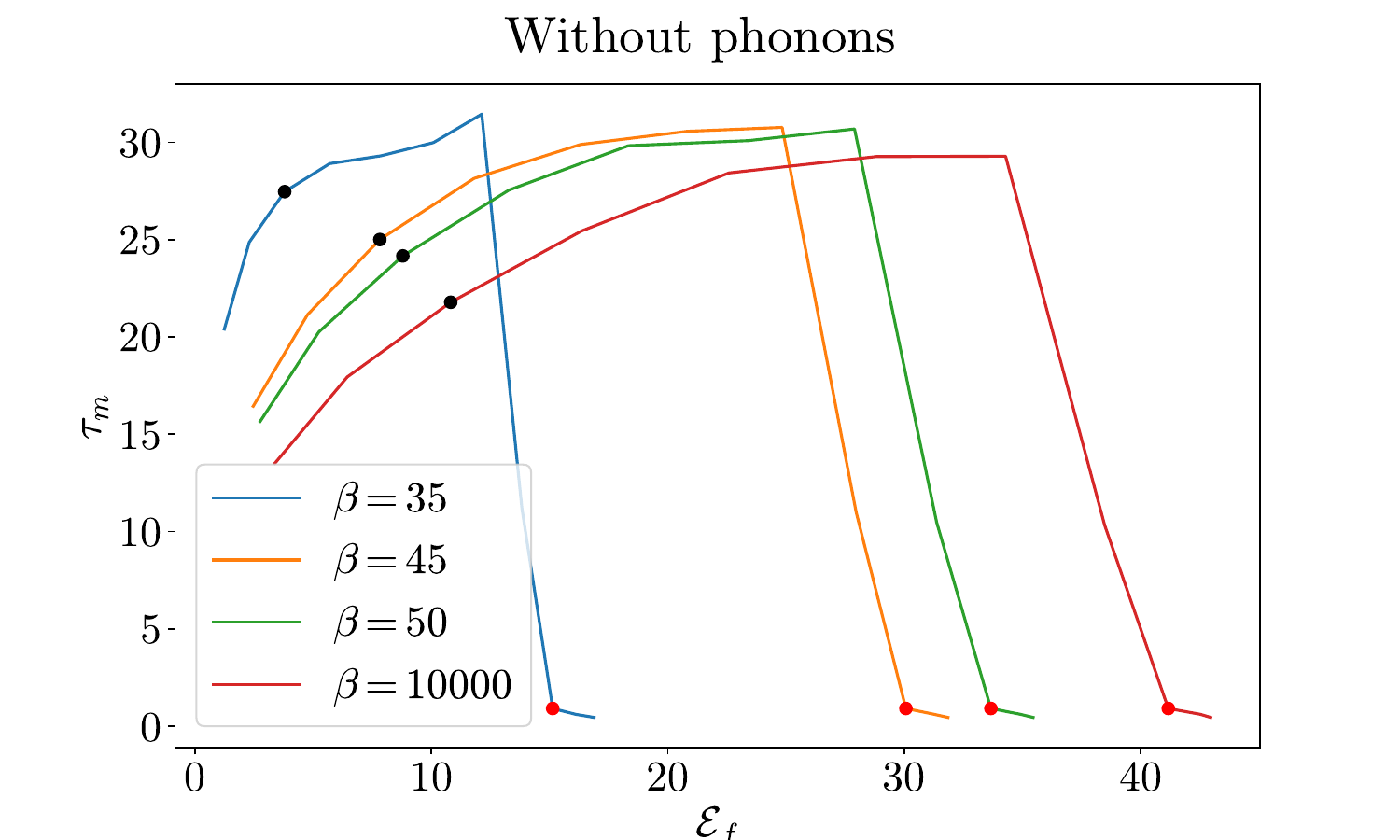}
	\caption{Melting time $\tau_m$ versus absorbed internal energy for different choices of inverse
	temperature $\beta = 1/T$. The black dots mark the melting time
	at the absorbed energy which equals the difference in internal energy of the
	superconducting and normal state at time zero, $\mathcal E_{\mr{sc}}$.
	The red dots mark the melting time at the absorbed energy where the average
	of the order parameter goes to zero at some point during the time evolution.
	Parameters used are $N_x = 100$ and $\omega = \gamma = 0$.
	}
	\label{fig:tauvsdeltae}
\end{figure}

\begin{figure}
	\centering
	\includegraphics[width=1\linewidth]{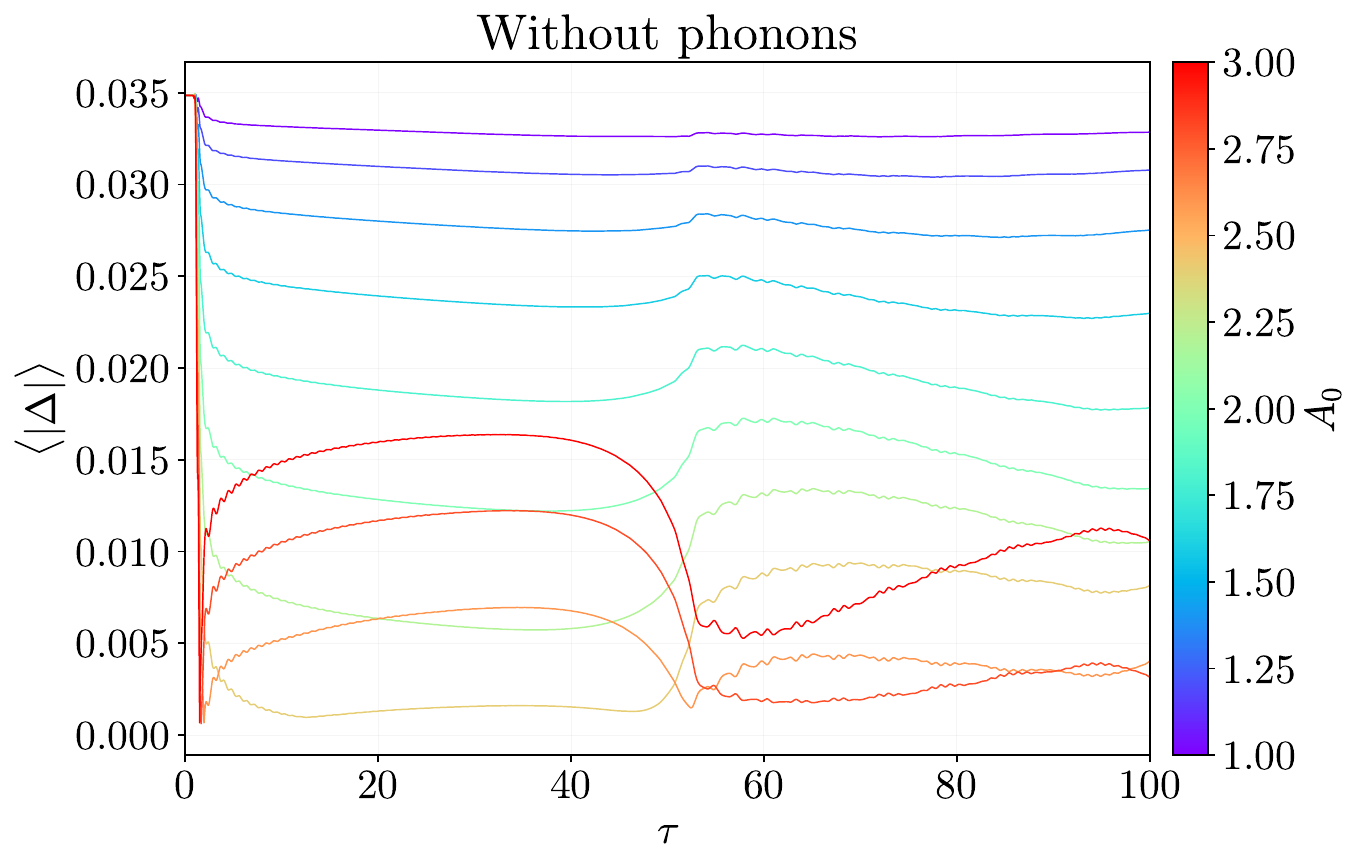}
	\caption{Several time evolutions of the order parameter averaged over sites at
	inverse temperature $\beta = 35$ with different choices of the parameter $A_0$
	varying between $1-3$. Parameters used are $N_x = 100$ and $\omega = \gamma = 0$.
	}
	\label{fig:rainbow}
\end{figure}

\textit{Results}. We start by considering the case without phonons. The temperature
dependence of $\tau_m$ is shown in Fig.~\ref{fig:tauvsdeltae}. Here the melting
time at each temperature is plotted versus the total absorbed energy after the laser
pulse has subsided. Although the absolute value of the absorbed energy is different
for each temperature, the qualitative behavior is similar in that the maximum of
$\tau_m$ always occurs for an absorbed energy between $\mathcal E_{\mr{sc}}$ and
$\mathcal E_d$, which is defined as the energy where the order parameter reaches
zero at some point during the time evolution. The black dots mark the values where
$\mathcal E_f = \mathcal E_{\mr{sc}}$ for each temperature, and the red dots mark
the values where $\mathcal E_f = \mathcal E_d$ respectively. The superconductivity
is destroyed completely for all temperatures shortly after the peak value of $\tau_m$.
This is in agreement with the recent experiment Ref. \cite{guillermo_arxiv_2026}.
Our model does not include any specific dissipation
mechanism, which means that the system should persist in a non-equilibrium state.
However, that does not mean that the order parameter cannot reach a stationary value
\cite{peronaci_prl_15}. For instance, oscillatory behavior of the order parameter
can be lost for large times despite the lack of dissipation in the system due to
so-called quench-induced decoherence \cite{schiro_prl_14}. In effect, we are assuming
that dissipation mechanisms of energy, such as heat loss to the environment, occur
at longer time scales than the $\tau$-values considered in Fig. \ref{fig:rainbow}.
The plot shows that the order parameter is gradually suppressed with increasing $A_0$,
until it finally changes sign around $A_0\simeq 2.5$. This causes the absolute value
to recover abruptly and results in a sharp reduction in melting time since the minimum
value of $\Delta$ turns into a local maximum of $|\Delta|$ as $\Delta$ changes sign.
The laser pulse thus causes an "overshoot" of the suppression of the order parameter
beyond zero into negative values, whereas the U(1) symmetry of the superconducting
condensate after the laser pulse has subsided renders $|\Delta|$ the physically
meaningful quantity. 
\begin{figure}
	\centering
	\includegraphics[width=1\linewidth]{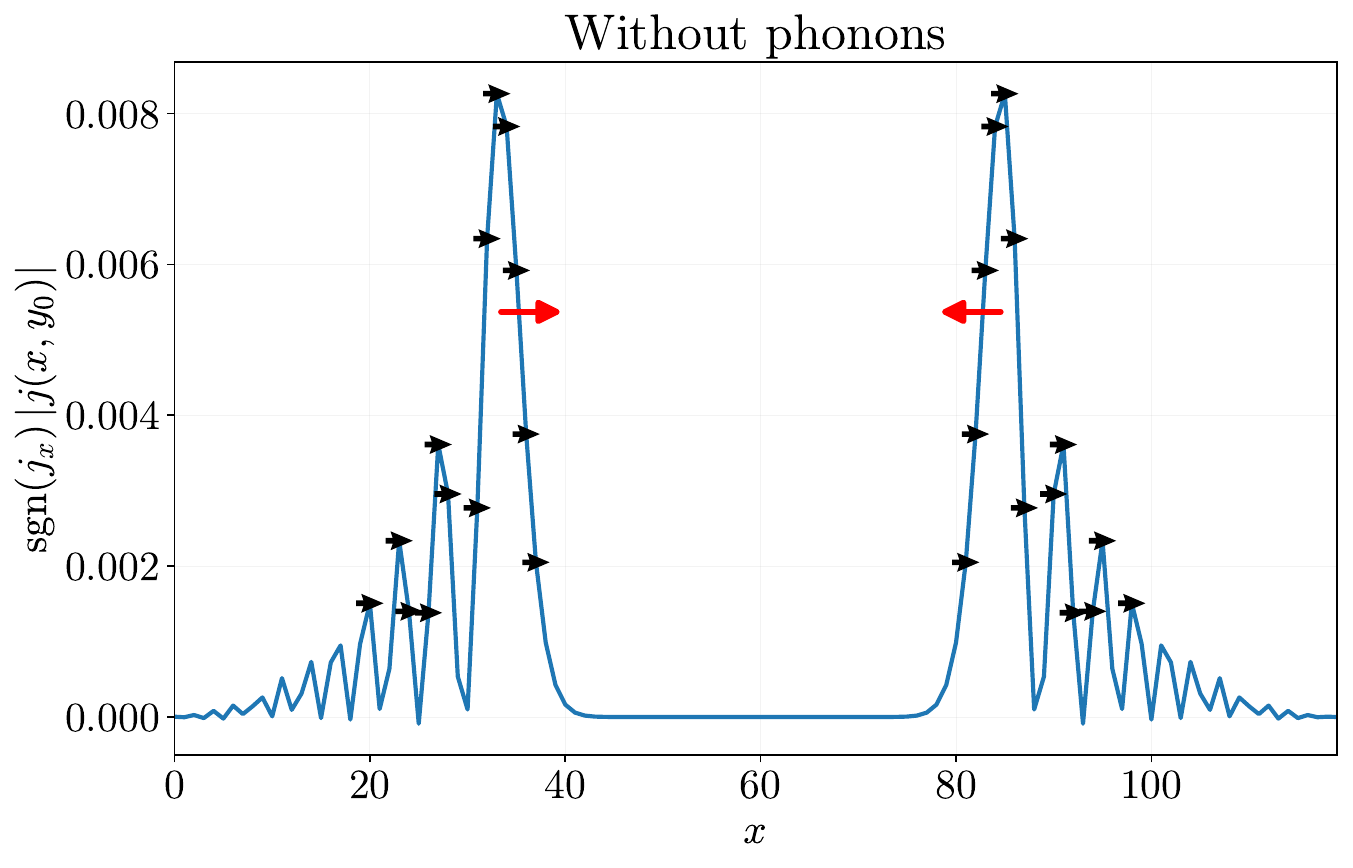}
	\includegraphics[width=1\linewidth]{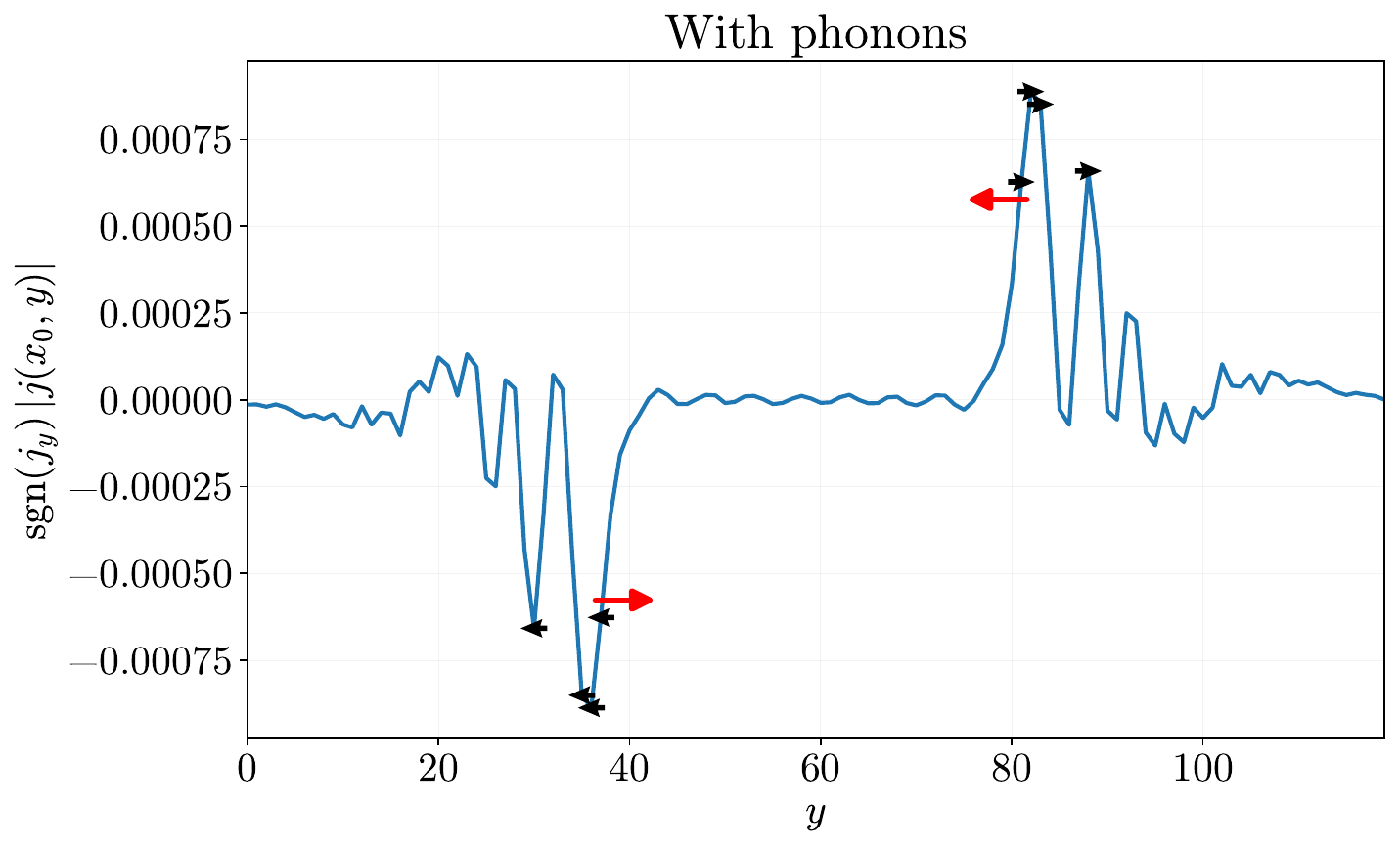}
	\caption{Direction and signed magnitude of currents of a slice of the superconductor at $\tau = 20$.
	The upper plot shows a slice $y_0 = 60$ for a superconductor without phonons.
	Both wavefronts are moving towards each other and the other end of the lattice as indicated by the red arrows,
	however the current is all in the $+x$-direction as indicated by the black arrrows.
	The lower plot shows a slice $x_0 = 60$ for a superconductor with phonons.
	Both wavefronts are moving towards each other but in this case the direction of the current
	is opposite that of the wavefront for both ends.
	Parameters used are $N_x = 120$, $A_0 = 2.2$, $\beta = 10000$ for both, and
	$\gamma = 0.05$, $\omega = 0.3$ for the case with phonons.
	}
	\label{fig:currents_a0_2p4_omega_0_beta_10000_t_5p5}
\end{figure}

\begin{figure}
	\centering
	\includegraphics[width=1\linewidth]{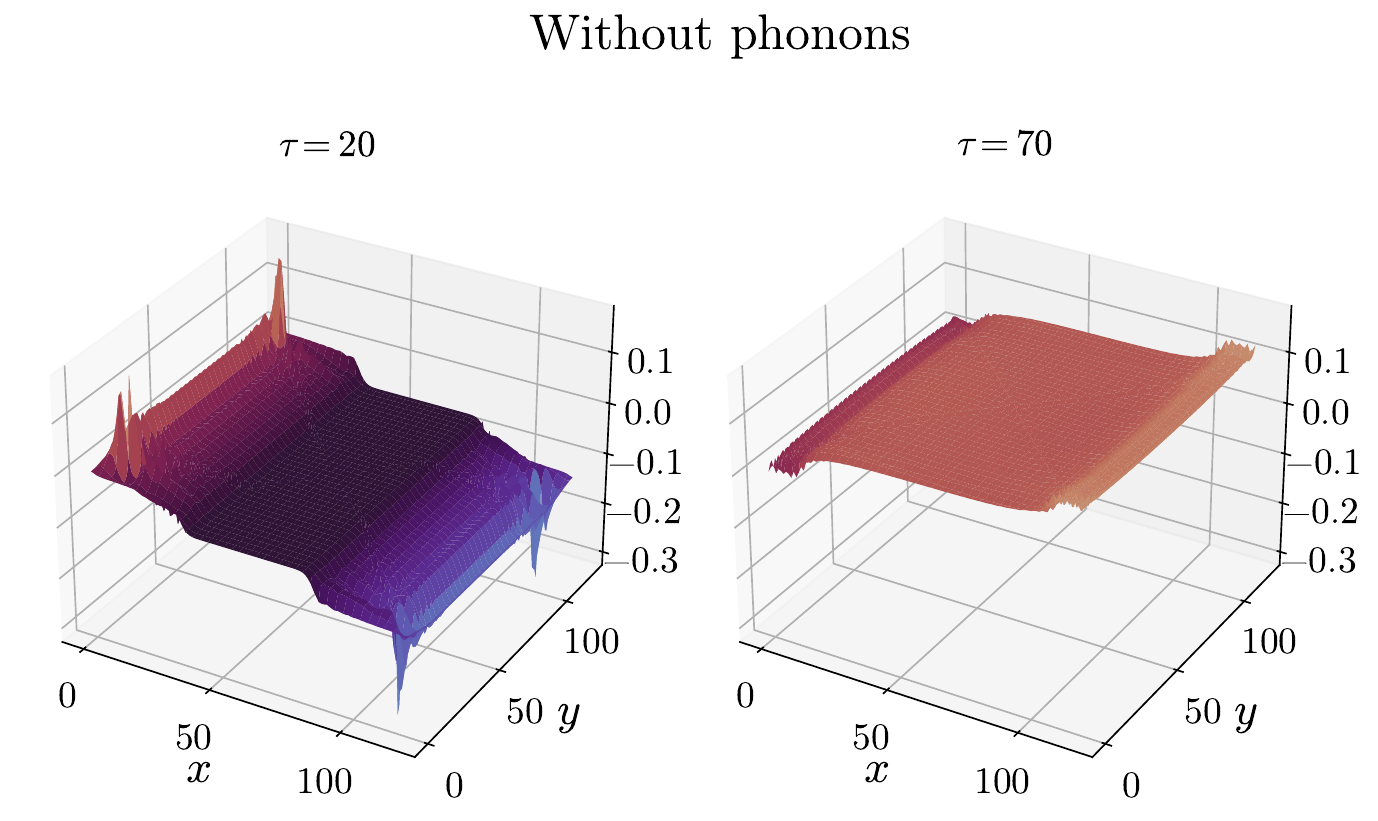}
	\caption{Spatially resolved phase of the order parameters,
	$\mathrm{arg}(\Delta_i)$ at two different times, showing the correspondence between
	the gradient of the phase and the currents.
	Parameters used are $N_x = 120$, $A_0 = 2.2$, $\gamma = \omega = 0$ and $\beta = 10000$.
	}
	\label{fig:phaseunwrap}
\end{figure}

The laser pulse induces transient bond-currents
\begin{equation}
	j_{i, i + \hat e} = - 2 \mathfrak{Im} [\mathrm{tr} \, (t_{i, i + \hat e} \, \rho_{i, i + \hat e})], 
\end{equation}
in the lattice, and spatial variations in the phase of the order parameters, $\phi_i$.
At low fluences, the magnitude of the current is approximately constant as a function
of $A_0$, and the spatial variation of the phase is small. However, after a critical
value of $A_0$ is exceeded, corresponding to the energy $\mathcal E_f$ required
to completely drive the order parameter to zero, the pulse induces strong spatial
variations in the phase, and the magnitude of the currents increases rapidly. 
The flow pattern of the currents is shown in Fig.~\ref{fig:currents_a0_2p4_omega_0_beta_10000_t_5p5}
and consists of one magnitude-wavefront of $+x$-direction currents traveling from
$x = 0$ towards $x = N_x$, and one front of $+x$-direction currents traveling the
opposite way from $x = N_x$ to $x = 0$, where the direction is chosen by the wavevector
of the vector potential. The wavefronts move with a speed of a little less than
two lattice-sites per $\tau$ (in units of $t$), consistent with the dispersion-relation
of a 2D tight-binding model. This leads to a inherent time-scale of the system,
which is the time needed for the two wavefronts to travel across the lattice and
reflect on the boundary, flipping the sign of the currents. This event can be seen
in the spatial variation of the phases $\phi_i$, where the gradient of the phase
switches sign at $\tau \approx N_x / 2$, see Fig.~\ref{fig:phaseunwrap}. To understand
why this is the case, imagine the laser pulse shifting the distribution of electrons
equally across the whole lattice back and forth in the $\pm x$-direction during
the pulse. After the pulse subsides, one of the edges has a surplus of electrons
whilst the other one has a deficit, and the middle of the lattice has shifted equally
and hence has neither surplus nor deficit. The edge with the deficit refills with
electrons, causing the current to move towards the left but leaving a deficit where
the electrons were, which in turn causes the magnitude-front to move in the other
direction as the distribution evens out. The opposite happens on the other side,
where the surplus of electrons at the edge starts to spill out, causing a new surplus
of electrons at the next site, which cascades into the current and magnitude-front
moving in the same direction. On a finite lattice, this is caused by the hard boundary
conditions. However, the hard boundary conditions are not necessary. The natural
"boundaries" caused by a vector potential varying in space is also sufficient to
cause regions with a surplus or deficiency of electrons. For example by assuming
a laser pulse modulated by a Gaussian in the $x$- and $y$-directions on a lattice
with periodic boundary conditions also creates current wavefronts with similar characteristics.
More details are provided in the supplemental material \cite{supplemental}. Whereas
pulse-induced currents have been discussed previously \cite{puviani_prb_23}, this
is to the best of our knowledge the first visualization of their actual flow pattern
in real space. Moreover, we underline that the current excitation predicted in this
work is different from a superfluid plasmon recently observed in \cite{hoegen_nature_26},
since such a plasmon relies on Coulomb interaction for its existence, which is not
included in our model.

\begin{figure}
	\centering
	\includegraphics[width=1\linewidth]{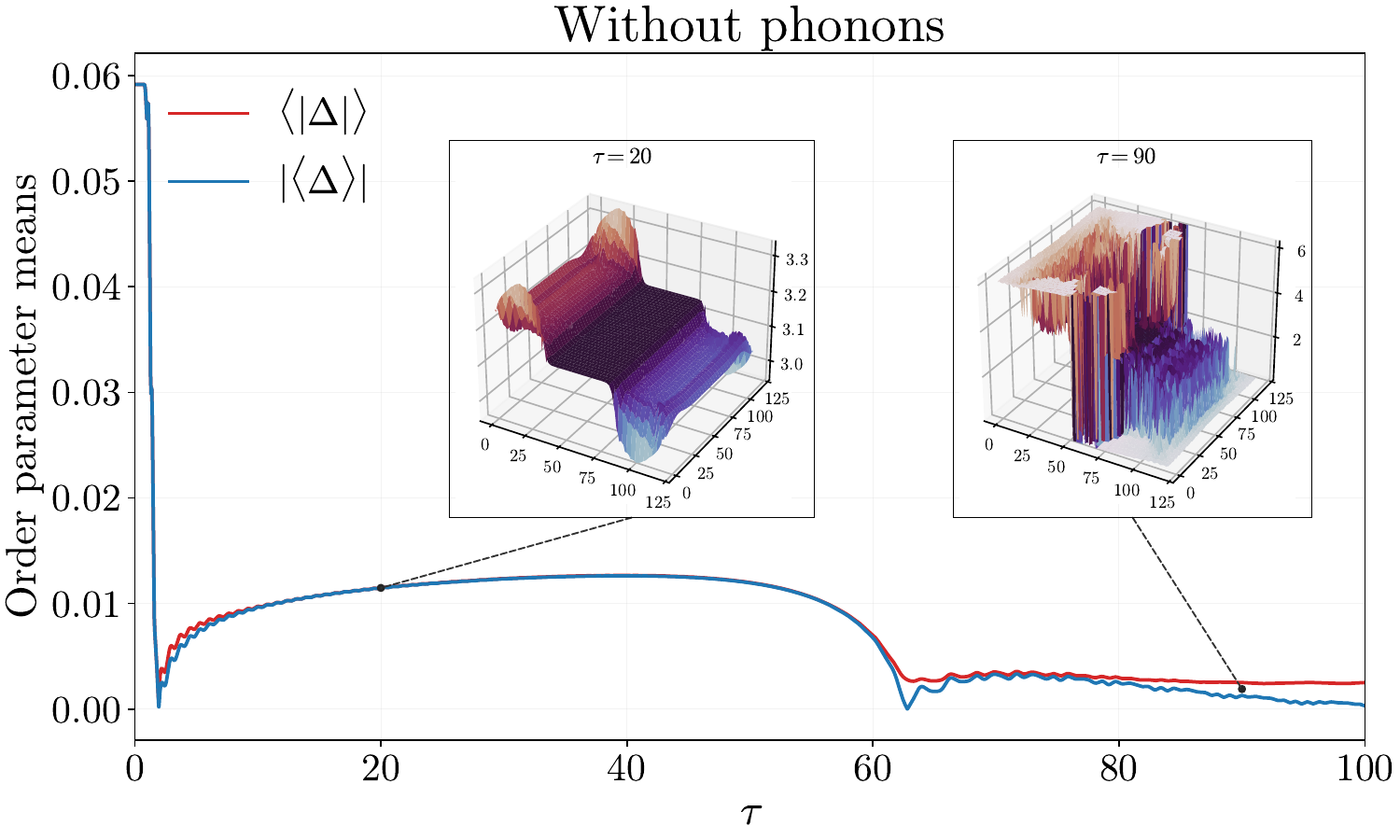}
	\caption{Difference between $\langle | \Delta | \rangle$ and $|\langle \Delta \rangle |$,
	showing how coherence of phases between different lattice sites are
	lost after a reflection at the boundaries leading to a suppression of the order
	parameter. Parameters used are $N_x = 120$, $A_0 = 2.6$, $\gamma = \omega = 0$ and $\beta = 10000$.
	}
	\label{fig:twomeans}
\end{figure}

The spatial distribution of the phase has an additional effect, most notably in cases
where the order parameter is nearly suppressed. In these situations, the phase
loses coherence across the lattice which leads to further suppression of the order
parameter. This may be seen in Fig.~\ref{fig:twomeans}, where both
$\langle |\Delta| \rangle \equiv 1 / N \sum_i |\Delta_i|$
and $|\langle \Delta \rangle | \equiv |1/N \sum_i \Delta_i |$ are shown. When the
coherence of the phases at different sites is lost, the phases on two different
sites is essentially uncorrelated. When the order parameter is averaged over all
sites, the random phases cancel out, leading to a reduction of the order parameter.
Thus, we find that a laser-pulse of sufficient magnitude destroys superconductivity
by means of inducing strong spatial phase fluctuations across the sample. 


We proceed to determine how the quench-induced dynamics of the superconducting order
parameter is altered when there exists a coupling between the superconducting electrons
and an optical phonon mode $\omega$. In this way, there can occur an energy exchange
between the superconducting condensate and phonons in the system. Our self-consistent
model thus captures both the extent to which phonons influence the melting of the
superconducting state and how efficient the energy transfer between the superconducting
condensate and the phonons is as a function of the injected energy through the laser
quench. As pointed out in Ref. \cite{demsar2003pair}, acoustic phonons may play a
key role in materials where optical phonon energies exceed $2\Delta$. Incorporation
of such modes into the present framework may thus be an interesting future venue
to explore.

The melting time and characteristics of the time evolution are detailed in the
supplemental material (SM) \cite{supplemental}, which includes videos of the spatially
resolved order parameter and current time dynamics. With phonons or with a non-zero
chemical potential, the time evolution renders the order parameter $\Delta$ to become
fundamentally complex, as opposed to the mostly real $\Delta$ obtained in the time
evolution without phonons, especially at low fluences.Specifically, the phonons induce a constantly
rotating phase $|\Delta_i|e^{i(\phi_i + \omega \tau)}$ reminiscent of a Goldstone
mode lingering after the pulse duration, in addition to a stronger spatial variation
of the phase $\phi_i$. Interestingly, the phonons modify the current flow in the
superconductor. When the optical phonon mode is present, currents induced by small
laser fluences move not only in the $x$-direction but also feature a weak current
with wavefronts travelling in the $y$-direction, as may be seen in Fig.~\ref{fig:currents_a0_2p4_omega_0_beta_10000_t_5p5}. 
In fact, it follows analytically from our equations of motion that electron-phonon
scattering physically has a very similar effect as a self-consistently determined
spatially varying chemical potential \cite{supplemental}, the latter also causing
electron scattering as it mimicks the presence of disorder.

\textit{Concluding remarks.} We have developed a microscopic real-space model of
laser pulse-induced melting of superconductivity, demonstrating that it reproduces
the experimentally observed critical slowing-down of the melting time near the condensation
energy. This goes substantially beyond previously employed phenomenological theories,
including the possibility to now treat ultrafast dynamics in heterostructures. Moreover,
we reveal that strong pulses generate phase decoherence and backward-wave-like current
textures with opposite phase and group velocity. This enables direct, experimentally
testable links between absorbed energy, phase disorder, and current flow. Our findings
are of relevance not only for nonequilibrium materials science, but also with respect
to the usage of superconducting dynamics in terahertz technology.

\begin{acknowledgments}
\textit{Acknowledgments.} K.B.H. and J. L. were supported by the Research Council
	of Norway through Grant No. 353894 and its Centres of
	Excellence funding scheme Grant No. 262633 “QuSpin.”
	Support from Sigma2 - the National Infrastructure for High
	Performance Computing and Data Storage in Norway, project
	NN9577K, is acknowledged.
\end{acknowledgments}

\bibliography{references}

\end{document}


\begin{center}
	\textbf{\Large Supplemental Material}
\end{center}

\section{Theory}
\subsection{Equations of motion}
The minimal set of time derivatives necessary to produce a closed system of equations
are, defining
$\rho_{ij}^{\sigma \sigma'} \equiv \expect{c_{i\sigma}^\dag c_{j \sigma'}}$,
$\Gamma_{ij}^{\sigma \sigma'} \equiv \expect{c_{i\sigma}^\dag c_{j\sigma'}^\dag}$,
$F_{ij}^{\sigma \sigma'} \equiv \expect{c_{i\sigma}c_{j\sigma'}}$,
$X_i \equiv \expect{b_i^\dag + b_i}$, $P_i \equiv \expect{b_i - b_i^\dag}$,
$n_i^b \equiv \expect{b_i^\dag b_i}$,
\begin{equation} \label{rho_deriv}
	\begin{split}
		 & \mathrm i\frac{d}{d\tau} \rho_{ij}^{\sigma \sigma} =
		\\ &\sum_k \bigg( (t_{jk} + t_{kj}^*) \rho_{ik}^{\sigma \sigma}
		- (t_{ki} + t_{ik}^*)\rho_{kj}^{\sigma \sigma} \bigg)
		\\ &+ \mathrm{sgn} \, \sigma \, (\Delta_j \Gamma_{ij}^{\sigma \bar\sigma}
		- \Delta_i^* F_{ij}^{\bar \sigma \sigma})
		\\ &+ \gamma(X_j - X_i )\rho_{ij}^{\sigma\sigma} 
		\\ &+ (\mu_j - \mu_i)\rho_{ij}^{\sigma \sigma},
	\end{split}
\end{equation}
where
\begin{align}
	\mathrm{sgn}\, \sigma & = \begin{cases}+1, \quad \sigma = \up \\
		                          -1, \quad \sigma = \dn\end{cases}  \\
	\bar\sigma            & = \begin{cases}\dn, \quad \sigma = \up \\
		                          \up, \quad \sigma = \dn\end{cases}.
\end{align}
Note that $F_{ij}^{\bar\sigma \sigma} = (\Gamma_{ji}^{\sigma \bar\sigma})^*$
and hence only one of $F$ or $\Gamma$ are required for the equations to close.

Similarily, for $\Gamma$, only $\Gamma^{\up \dn}$ is required for
the complete set,
\begin{equation} \label{gamma_deriv}
	\begin{split}
		 & \mathrm i \frac{d}{d\tau} \Gamma_{ij}^{\up \dn} =
		\\ & -\sum_k \bigg( (t_{ki} + t_{ik}^*)\Gamma_{kj}^{\up \dn}
		+ (t_{kj} + t_{jk}^*)\Gamma_{ik}^{\up\dn} \bigg)
		\\ &+ \Delta_j^* \rho_{ij}^{\up\up}
		- \Delta_i^* (\delta_{ij} - \rho_{ji}^{\dn\dn})
		\\ &- \gamma(X_i + X_j)\Gamma_{ij}^{\up\dn}
		\\ &- (\mu_i + \mu_j)\Gamma_{ij}^{\up\dn}.
	\end{split}
\end{equation}

For the phonon part, the equations of motion are given by
\begin{align}
	 & \mathrm i \frac{d}{d\tau} X_i =
	\omega_{\text{ph}}P_i, 
    \label{x_deriv}
	\\ &\mathrm i \frac{d}{d\tau} P_i = \omega_{\text{ph}} X_i
	+ 2 \gamma \expect{n_i^f}, 
    \label{p_deriv}
	\\ &\mathrm i \frac{d}{d\tau}\expect{n_i^b} =
	-\gamma P_i\expect{n_i^f}, \label{occ_deriv}
\end{align}
where Eq. \eqref{occ_deriv} is only needed for the purposes of calculating
the internal energy.

\subsection{Initial conditions}
The laser pulse is turned on at time $t = 0$, after which the time dynamics is
calculated according to the Heisenberg equations for the various correlators
of the system.  In order to increase the numerical accuracy of the results,
a precise set of initial conditions for the correlations are needed. This is
accomplished by performing a mean-field approximation on the phonons under the
assumption that before the laser pulse, the system is in equilibrium and at a
very low temperature, hence using $b_i^\dag b_i \rightarrow f^b(\omega_{\mr{ph}})$,
where $f^b$ is the Bose-Einstein distribution, $b_i^\dag + b_i \rightarrow X_i(0)$
and Eq.\eqref{p_deriv},
\begin{equation}
	\begin{split}
		 & \mathrm i \frac{d}{d\tau} P_i = 0 = \omega_{\text{ph}} X_i
		+ 2\gamma \expect{n_i}
		\\ &\implies X_i(-\infty)
		= -\frac{2\gamma}{\omega_{\text{ph}}} \expect{n_i}.
	\end{split}
\end{equation}
This enables us to write the Hamiltonian in standard BdG form and find a
reasonable self-consistent initial condition for the time evolution.
\begin{equation}
	\begin{split}
		 & \hat H(-\infty) =
		\\ &\sum_{ij\sigma}\frac{1}{2}(
		c_{i\sigma}^\dag t_{ij}c_{j\sigma} - c_{i\sigma}t_{ji}c_{j\sigma}^\dag)
		\\ &+ \sum_i \frac{1}{2} [ \Delta_i (c_{i\up}^\dag c_{i \dn}^\dag
			- c_{i\dn}^\dag c_{i\up}^\dag)
			+ \Delta_i^* (c_{i\dn}c_{i\up} - c_{i \up} c_{i \dn}) ]
		\\ &+ \sum_{i\sigma} \frac{1}{2} [c_{i\sigma}^\dag (\mu_i + X_i(0)) c_{i\sigma}
			- c_{i\sigma}(\mu_i + X_i(0)) c_{i \sigma}^\dag]
		\\ &+ \sum_i \omega_{\mr{ph}} f^b(\omega_{\mr{ph}}).
	\end{split}
\end{equation}

Introducing
$\hat c_i \equiv (c_{i\up}, c_{i\dn}, c_{i\up}^\dag, c_{i\dn}^\dag)^T$ allows
us to write the Hamiltonian in the form
\begin{equation}
	\hat H(-\infty) = E_0 + \frac{1}{2} \sum_{ij}\hat c_{i}^\dag \hat H_{ij} \hat c_{j}
	= E_0 + \frac{1}{2}\check c^\dag \check H \check c,
\end{equation}
where
$E_0 = \omega_{\mr{ph}}N_{\mr{ph}} f^b(\omega_{\mr{ph}}) + \sum_i |\Delta_i|^2 / U$
and
\begin{equation}
	\hat H_{ij} = \begin{pmatrix}
		(X_i(0) + \mu_i)\delta_{ij} + t_{ij}  & \Delta_i (- \mr i \sigma_y)\delta_{ij} \\
		\Delta_i^*(\mr i \sigma_y)\delta_{ij} & -(X_i(0) + \mu_i)\delta_{ij} - t_{ji}.
	\end{pmatrix}
\end{equation}
We solve the Hamiltonian self-consistently by making a guess for the values of
$\Delta_i$ and $X_i(0)$ and diagonalize $\check H$. The values for $\Delta_i$
and $X_i(0)$ on the next iteration are obtained from the eigenvectors
$\check \chi_n$ with components
$(\check \chi_n)_i = (u_{n, i\up}, u_{n, i\dn}, v_{n, i\up}, v_{n, i\dn})^T$ and
the relations
\begin{align}
	c_{i\sigma} & = \sum_{ n > 0 }
	(u_{n, i\sigma}\gamma_n + v_{n, i\sigma}^*\gamma_n^\dag)
	\\ \expect{\gamma_n \gamma_m^\dag} &= (1 - f(E_n))\delta_{nm}
	\\ \expect{\gamma_n^\dag \gamma_m} &= f(E_n)\delta_{nm},
\end{align}
yielding
\begin{equation}
	\begin{split}
		\Delta_i = U \sum_{n > 0} \bigg( u_{n,i\dn} (1 - f(E_n)) v_{n, i\up}^*
		\\ + u_{n,i \up} f(E_n) v_{n, i\dn}^* \bigg),
	\end{split}
\end{equation}
and
\begin{equation}
	\begin{split}
		\expect{n_i} =
		\sum_{\sigma, n>0}\bigg( |u_{n, i\sigma}|^2 f(E_n)
		\\ + |v_{n,i\sigma}|^2 (1 - f(E_n))\bigg),
	\end{split}
\end{equation}
where $\sum_{n>0}$ is short hand for $\sum_{\{n \, : \, E_n > 0 \}}$ and
$f$ is the Fermi-Dirac distribution.

Once the Hamiltonian has converged to within an error margin, the correlations
of the last iteration are used to serve as the initial condition for the time
evolution via
\begin{equation}
	\begin{split}
		\expect{c_{i\dn}c_{j\up}} =
		\sum_{n>0} \bigg( u_{n, i\dn}(1 - f(E_n)) v_{n, i\up}^*
		\\ + v_{n, i\dn}^* f(E_n) u_{n, j\up} \bigg),
	\end{split}
\end{equation}
and
\begin{equation}
	\begin{split}
		\expect{c_{i\sigma}^\dag c_{j\sigma}} =
		u_{n,i\sigma}^*f(E_n)u_{n,j\sigma}
		\\ + v_{n,i\sigma}(1-f(E_n))v_{n,j\sigma}^*.
	\end{split}
\end{equation}

\subsection{Critical slowing down}

The critical slowing of the time-dynamics as the order parameter
approaches zero can also be illustrated heuristically using
time-dependent Ginzburg-Landau (TDGL) theory as follows. For an order parameter
$\phi$ with free-energy density
\begin{equation}
	f(\phi) = \frac a 2 \phi^2 + \frac b 4 \phi^4,
\end{equation}
where $b > 0$ and $a = a_0(T - T_c)$, the TDGL equation
is given by, in the case of homogeneous order parameter,
\begin{equation}
	\partial_\tau \phi = - \Gamma \frac{d f}{d \phi} = -\Gamma(a\phi + b\phi^3),
\end{equation}
where $\Gamma$ is a kinetic coefficient of the order parameter \cite{hohenberg_revmodphys_1977, goldenfeld_lectures_1992}
Linearizing around a stationary solution, we can expand around
$\phi(\tau) = \phi_0 + \delta \phi(\tau)$ giving,
\begin{equation}
\begin{split}
	&f'(\phi_0 + \delta \phi) = f'(\phi_0) + f''(\phi_0)\delta \phi + \mathcal O(\delta \phi^2) \\
	&= f''(\phi_0)\delta \phi + \mathcal O(\delta \phi^2),
\end{split}
\end{equation}
since stationarity implies $f'(\phi_0) = 0$. This yields a linearized equation,
\begin{equation}
	\partial_\tau (\delta \phi) = -\Gamma f''(\phi_0)\delta \phi.
\end{equation}
The solution to this equation is given by
\begin{equation}
	\delta \phi(\tau) = \delta \phi(0) e^{-\Gamma f''(\phi_0)\tau},
\end{equation}
with relaxation time $\tau_{\mr{rel}} = (\Gamma f''(\phi_0))^{-1}$.
The stationary solutions satisfy $f'(\phi_0) = 0$, giving
\begin{align}
	\phi_0(a + b\phi_0^2) = 0 \implies \phi_0 =
	\begin{cases}
		0, \quad a > 0 \\
		\pm \sqrt{-a/b}, \quad a < 0.
	\end{cases}
\end{align}
We use the free-energy density to evaluate the curvature as $T \rightarrow T_c$,
\begin{equation}
	f''(\phi_0) = a + 3b\phi_0^2 =
	\begin{cases}
		a, \quad T > T_c \\
		-2a, \quad T < T_c.
	\end{cases}
\end{equation}
In both cases, $f''(\phi_0) \rightarrow 0$ as $T \rightarrow T_c$ and hence the
relaxation time $\tau_{\mr{rel}}$ diverges. Thus, as the stationary order parameter
$\phi_0 \rightarrow 0$ when $T \rightarrow T_c$, the curvature $f''(\phi_0)$
vanishes and the decay time-scale grows, consistent with the increase of the melting time.

\subsection{Doubled Hamiltonian}
When the Hamiltonian does not contain any spin-mixing terms, \textit{e.g.} just a
straight hopping plus pairing Hamiltonian, the BdG Hamiltonian takes the form
\begin{equation}
	\hat H_{ij} = \begin{pmatrix}
		t_{ij}      & 0          & 0        & -\Delta_i \\
		0           & t_{ij}     & \Delta_i & 0         \\
		0           & \Delta_i^* & -t_{ji}  & 0         \\
		-\Delta_i^* & 0          & 0        & -t_{ij}
	\end{pmatrix}
\end{equation}
we can instead use the basis $(c_{i\up}, c_{i\dn}^\dag, c_{i\dn}, c_{i\up}^\dag)$
and rewrite it as
\begin{equation}
	\hat H_{ij} = \begin{pmatrix}
		t_{ij}     & \Delta_i & 0           & 0         \\
		\Delta_i^* & -t_{ji}  & 0           & 0         \\
		0          & 0        & t_{ij}      & -\Delta_i \\
		0          & 0        & -\Delta_i^* & -t_{ji}
	\end{pmatrix} = \begin{pmatrix}
		\hat H_{ij}^+ & 0             \\
		0             & \hat H_{ij}^-
	\end{pmatrix},
\end{equation}
which means that the set of eigenvalues for the full matrix are given by
$\{\pm E_1, \pm E_1, \pm E_2, \pm E_2, \dots, \pm E_{N}, \pm E_{N}\}$.
This is proven as follows. Assume $(\zeta_n)_i = (u_{i\up}, v_{i\dn})$
is an eigenvector of $\hat H^+_{ij}$ such that
\begin{equation}
	\begin{split}
		\sum_j \begin{pmatrix}
			       t_{ij}     & \Delta_i \\
			       \Delta_i^* & -t_{ji}
		       \end{pmatrix} \begin{pmatrix} u_{j\up} \\ v_{j\dn} \end{pmatrix}
		= E_n \begin{pmatrix} u_{i\up} \\ v_{i\dn} \end{pmatrix}.
	\end{split}
\end{equation}
Then
\begin{align}
	\sum_j t_{ij} u_{i\up} + \Delta_i v_{i\dn} = E_n u_{i\up} \\
	-\Delta_i^* u_{i\up} + \sum_j t_{ji} v_{j\dn} = E_n(-v_{i\dn})
\end{align}
This is written compactly:
\begin{equation}
	\sum_j \begin{pmatrix}
		t_{ij}      & -\Delta_i \\
		-\Delta_i^* & -t_{ji}
	\end{pmatrix} \begin{pmatrix} u_{i\up} \\ -v_{i\dn} \end{pmatrix}
	= E_n \begin{pmatrix}
		u_{i\up} \\ -v_{i\dn}
	\end{pmatrix}.
\end{equation}
Hence, if $(u_{i\up}, v_{i\dn})$ is an eigenvector of the upper block with
eigenvalue $E_n$, then $(u_{i\up}, -v_{i\dn})$ is an eigenvector of the lower
block with the same eigenvalue $E_n$. In a similar way, we can show that if
$(u_{i\up}, v_{i\dn})$ is an eigenvector with eigenvalue $E_n$, then
$(v_{i\dn}^*, u_{i\up}^*)$ is an eigenvector of the second block with eigenvalue
$-E_n$. There is also a second eigenvector of the upper block
$(v^*_{i\dn}, -u^*_{i\up})$ with eigenvalue $-E_n$. In total we can therefore
find the components of four eigenvectors
from the components of one.

Translated into the $4N$ case, we know that if there is an eigenvector
$(u_{i\up}, 0, 0, v_{i\dn})$ then there is one with the same eigenvalue given by
$(0, u_{i\up}, -v_{i\dn}, 0)$, one with negative eigenvalue at
$(v^*_{i\dn}, 0, 0, -u^*_{i\up})$ and one with negative eigenvalue
at $(0, v^*_{i\dn}, u^*_{i\up}, 0)$. Note that this means that we can drop the
spin indices.

Instead of working with the full $4N \times 4N$ matrix, we can thus
work with the $2N \times 2N$ matrix $\check H^+$ with components in the basis
$(c_{i\up}, c_{i\dn}^\dag)$
\begin{equation}
	\hat H_{ij}^+ = \begin{pmatrix}
		t_{ij}     & \Delta_i \\
		\Delta_i^* & -t_{ji}
	\end{pmatrix},
\end{equation}
and eigenvalues $\{E_1, E_2, \dots, E_{2N}\}$. This procedure saves computational
time. The eigenvalues are related by
\begin{equation}
	E_{n+N} = -E_n.
\end{equation}
For example, the anomalous correlator is given by
\begin{equation}
	\langle c_{i\dn} c_{i\up} \rangle_{2N} = \sum_{n=1}^{2N} \zeta_{n,i,1}^*\zeta_{n, i, 0} f(E_n).
\end{equation}
This is provably the same result as for the $4N$ case as can be seen by
starting from the full expression
\begin{equation}
	\begin{split}
		\langle c_{i\dn} c_{i \up} \rangle_{4N} = \sum_{n = 1}^{2N} \bigg(
		\check \chi_{n, i, 1} (1 - f(E_n)) \check \chi_{n, i, 2}^* \\
		+ \check \chi_{n, i, 3}^* f(E_n) \check \chi_{n,i,0} \bigg).
	\end{split}
\end{equation}
We then use what was shown above, namely if $(u_{ni}, 0, 0, v_{ni})$
is an eigenvector with eigenvalue $E_n$ then $(0, u_{ni}, -v_{ni}, 0)$ is also
an eigenvector $(\check \chi_{n+N})_i$ with eigenvalue $E_n$, to rewrite
the sum into a sum over $N$ positive eigenvalues via,
\begin{equation}
	\begin{split}
		\sum_{n = 1}^{N}\bigg(\check\chi_{n,i,1}(1-f(E_{n}))\check\chi_{n,i,2}^* + \check\chi_{n,i,3}^*f(E_{n})\check\chi_{n,i,0} \\
		+ \check\chi_{n+N,i,1}(1 - f(E_{n}))\check\chi_{n+N,i,2}^* + \check\chi_{n+N,i,3}^*f(E_n)\check\chi_{n+N,i,0}\bigg)       \\
		= \sum_{n=1}^N \bigg( 0 \cdot f(-E_{n})\cdot 0  + u_{ni} f(E_n) v_{ni}^*                                                  \\
		+ u_{ni}f(-E_n)(-v_{ni}^*) + 0 \cdot f(E_n) \cdot 0  \bigg)                                                               \\
		= \sum_{n=1}^N u_{ni}f(E_n)v_{ni}^* - u_{ni}(1 - f(E_n))v_{ni}^*
	\end{split}
\end{equation}
Hence this expression reduces to the same as the $2N \times 2N$ one
\begin{equation}
	\begin{split}
		\langle c_{i\dn} c_{i\up} \rangle_{2N} = \sum_{n=1}^{2N}\zeta_{n,i,1}^* \zeta_{n,i,0}f(E_n)   \\
		= \sum_{n = 1}^N \zeta_{n,i,1}^*f(E_n)\zeta_{n,i,0} + (-\zeta_{n,i,0}) \zeta_{n,i,1}^*f(-E_n) \\
		= \sum_{n = 1}^N u_{ni} f(E_n) v_{ni}^* - u_{ni}(1-f(E_n))v_{ni}^*.
	\end{split}
\end{equation}
\section{Results}

\subsection{Without phonons}

\begin{figure*}[p]
	\centering
	\includegraphics[width=1\linewidth]{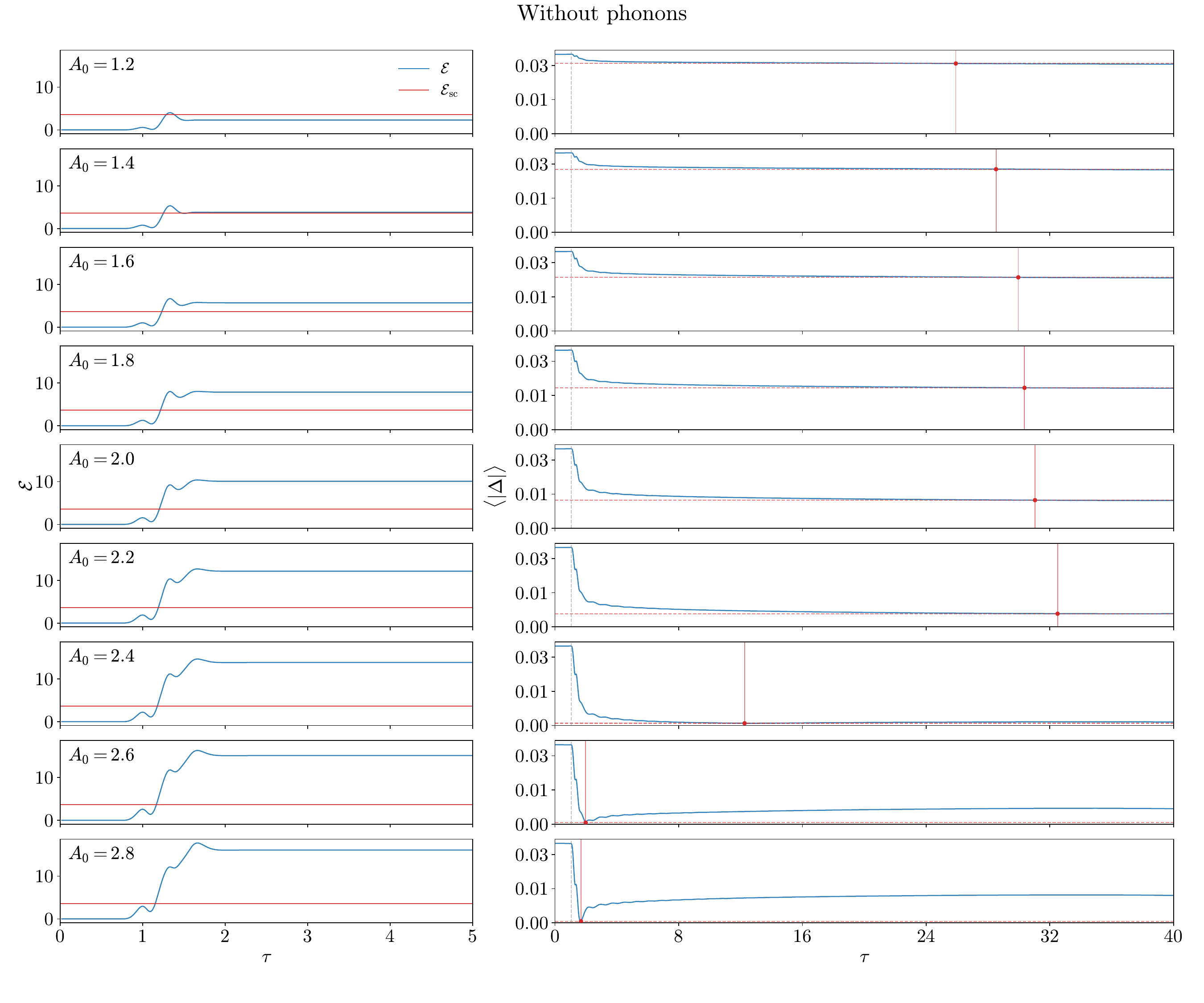}
	\caption{Time dynamics of the absorbed internal energy and the order parameter
	averaged over all sites. The leftmost plots shows the absorbed internal energy
	$\mathcal E$ in blue with gradually increasing laser amplitude $A_0$, with
	the horizontal red line marking the difference in internal
	energy between the superconducting and normal state before the laser pulse is
	applied. The rightmost plots shows the time evolution of the site-averaged order
	parameter $\expect{|\Delta|}$ in blue. The red dashed horizontal line marks
	the minimal value that $\expect{|\Delta|}$ attains during the time-evolution,
	and the solid red vertical line and dot shows the time $\tau_m$ where this minimum
	value is attained. Parameters used are $U = -1$, $\beta = 35$ and $\omega = \gamma = 0$.
	}
	\label{fig:deltatime}
\end{figure*}
The results of the time dynamics is shown for $\beta = 35$ in Fig.~\ref{fig:deltatime},
where the plot shows the time evolution of the change in internal energy, $\mathcal E(\tau)$,
and a site-averaged gap $\expect{|\Delta|} \equiv \sum_{i = 1}^N \frac1N |\Delta_i|(\tau)$.
The change in internal energy reaches a constant value some time after the laser
pulse has subsided, which we refer to as $\mathcal E_f$.
From top to bottom, the plots show the effects of increasing the energy absorbed
from the laser pulse.
The red horizontal line in the left column corresponds to the difference in
internal energy between the superconducting state and the normal state before the pulse
$\mathcal E_{\mr{sc}}$.
The blue line shows the change in internal energy $\mathcal{E}(\tau)$.
The right column shows the melting of the superconductivity after the applied
laser pulse. The red vertical line shows the time where the order parameter reaches
its minimal value, corresponding to $\tau_m$. The final absorbed energy initially
increases with laser amplitude $A_0$, with an accompanying increase in the
melting time. This changes at a critical value of $A_0$ where the melting time
starts decreasing rapidly. The value of $\mathcal E_f$ where the maximum of
$\tau_m$ occurs is temperature dependent, but it always occurs after
$\mathcal E_f$ surpasses $\mathcal E_{\mr{sc}}$, yet before the value of
$\mathcal E_f$ which is sufficient to completely destroy the superconductivity
and bring the order parameter to zero, $\mathcal E_d$.

The plot shows that the order parameter is gradually suppressed with increasing
$A_0$, until it finally changes sign around $A_0\simeq 2.5$. This causes the absolute
value to recover abruptly and results in a sharp reduction in melting time since
the minimum value of $|\Delta|$ turns into a local maximum as $\Delta$ changes sign.
The laser pulse thus causes an "overshoot" of the suppression of the order parameter
beyond zero into negative values, whereas the U(1) symmetry of the superconducting
condensate after the laser pulse has subsided renders $|\Delta|$ the physically
meaningful quantity. 

\subsection{With phonons}

We proceed to determine how the quench-induced dynamics of the superconducting order
parameter is altered when there exists a coupling between the superconducting electron
and an optical phonon mode $\omega$. In this way, there can occur an energy exchange
between the superconducting condensate and phonons in the system. Our self-consistent
model thus captures (i) to what extent the phonons influence the melting of the
superconducting state and (ii) how efficient the energy transfer between the superconducting
condensate and the phonons is as a function of the injected energy through the laser-quench.

\begin{figure}
	\centering
	\includegraphics[width=1\linewidth]{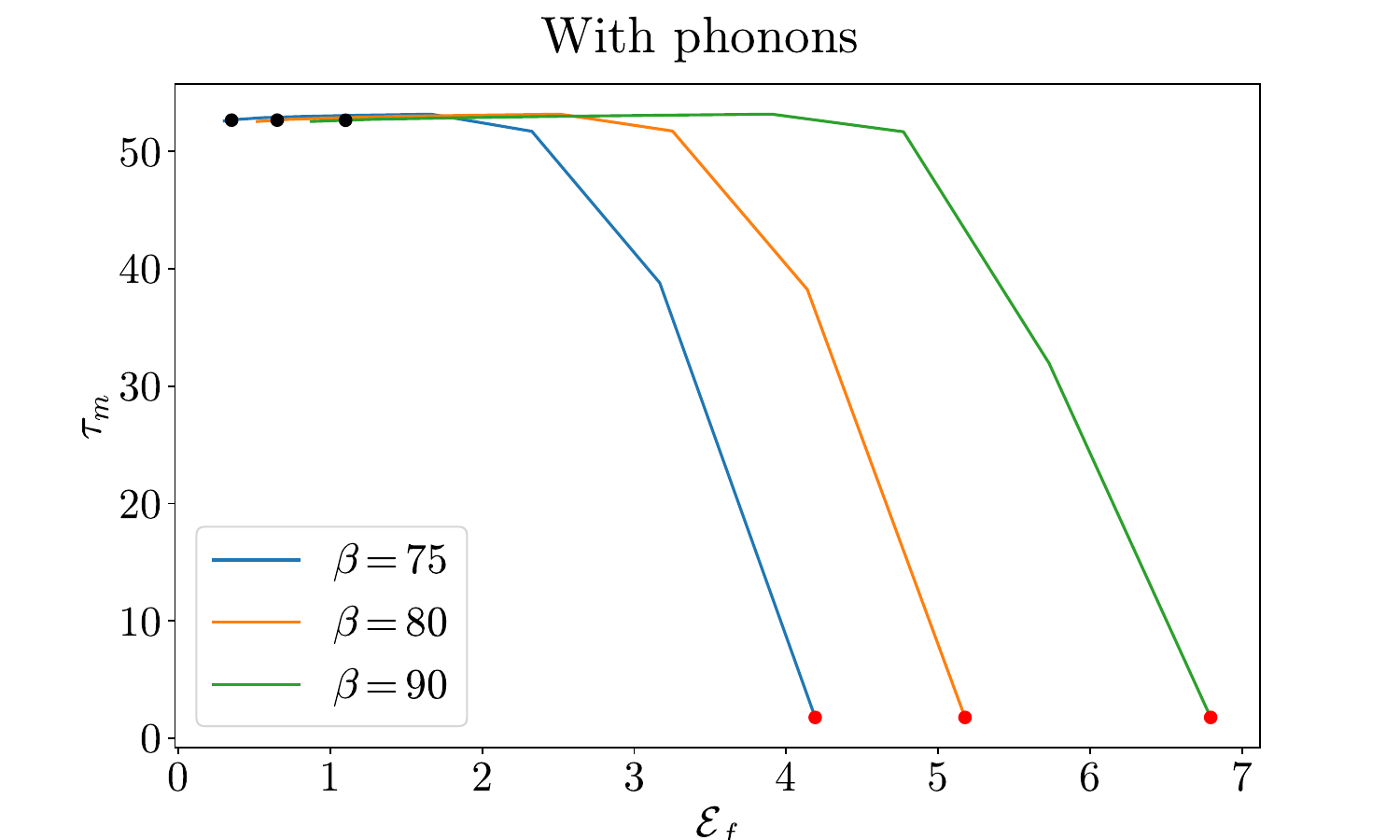}
	\caption{Melting time $\tau_m$ versus absorbed internal energy for different
	choices of inverse temperature $\beta = 1/T$ with phonons. The black dots mark
	the melting time at the absorbed energy which equals the difference in internal
	energy of the superconducting and normal state at time zero, $\mathcal E_{\mr{sc}}$.
	The red dots mark the melting time at the absorbed energy where the average
	of the order parameter goes to zero at some point during the time evolution. 
	Parameters used are $N_x = 100$, $\omega = 0.3$, and $\gamma = 0.05$.
	}
	\label{fig:tauvsdeltae_phonon}
\end{figure}
\begin{figure}
	\centering
    \includegraphics[width=0.95\linewidth]{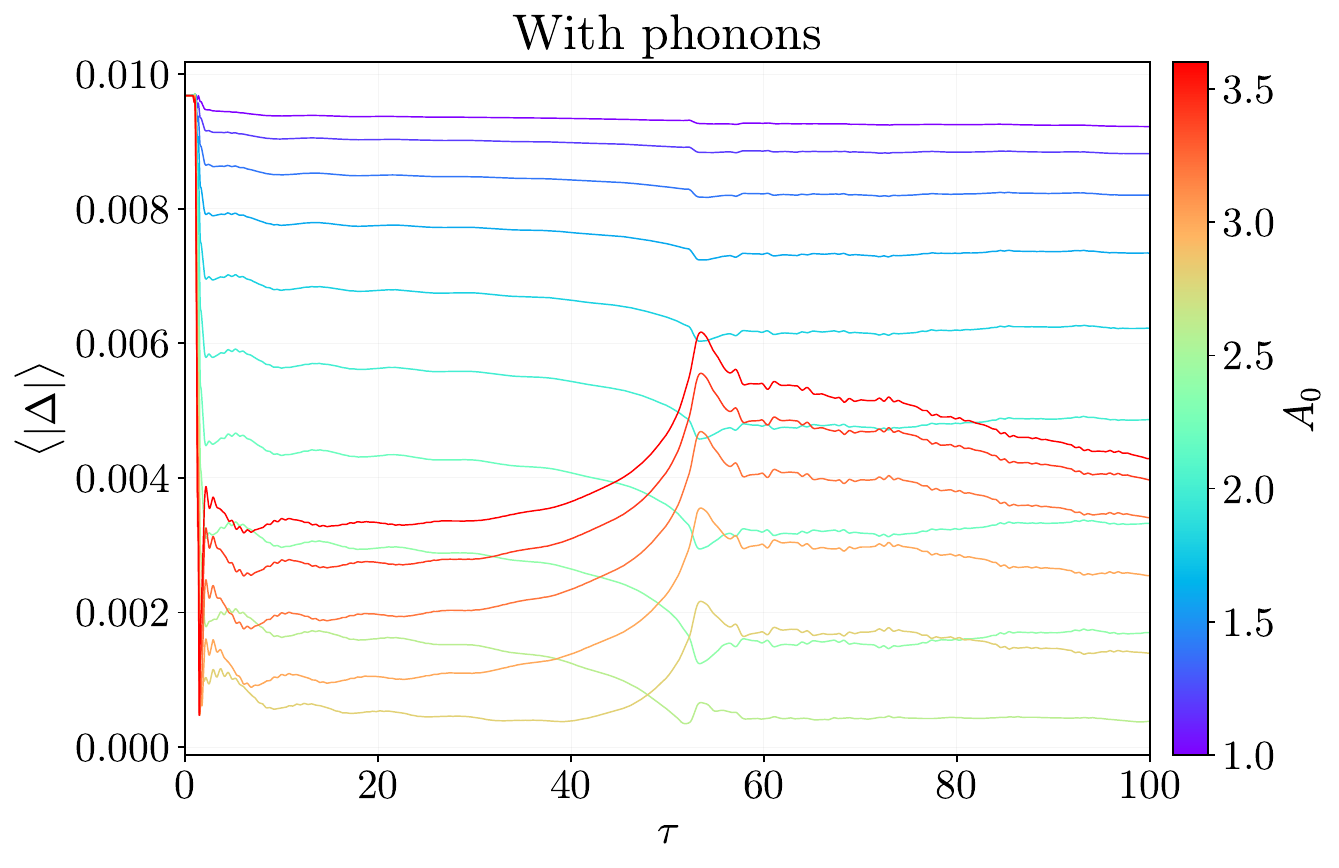}
	\caption{Several time evolutions of the order parameter averaged over sites at
	inverse temperature $\beta = 75$ with different choices of the laser amplitude parameter $A_0$.
	Parameters used are $N_x = 100$, $\omega = 0.3$ and $\gamma = 0.05$.
	}
	\label{fig:rainbow_phonon}
\end{figure}

\begin{figure*}[p]
	\centering
	\includegraphics[width=1\linewidth]{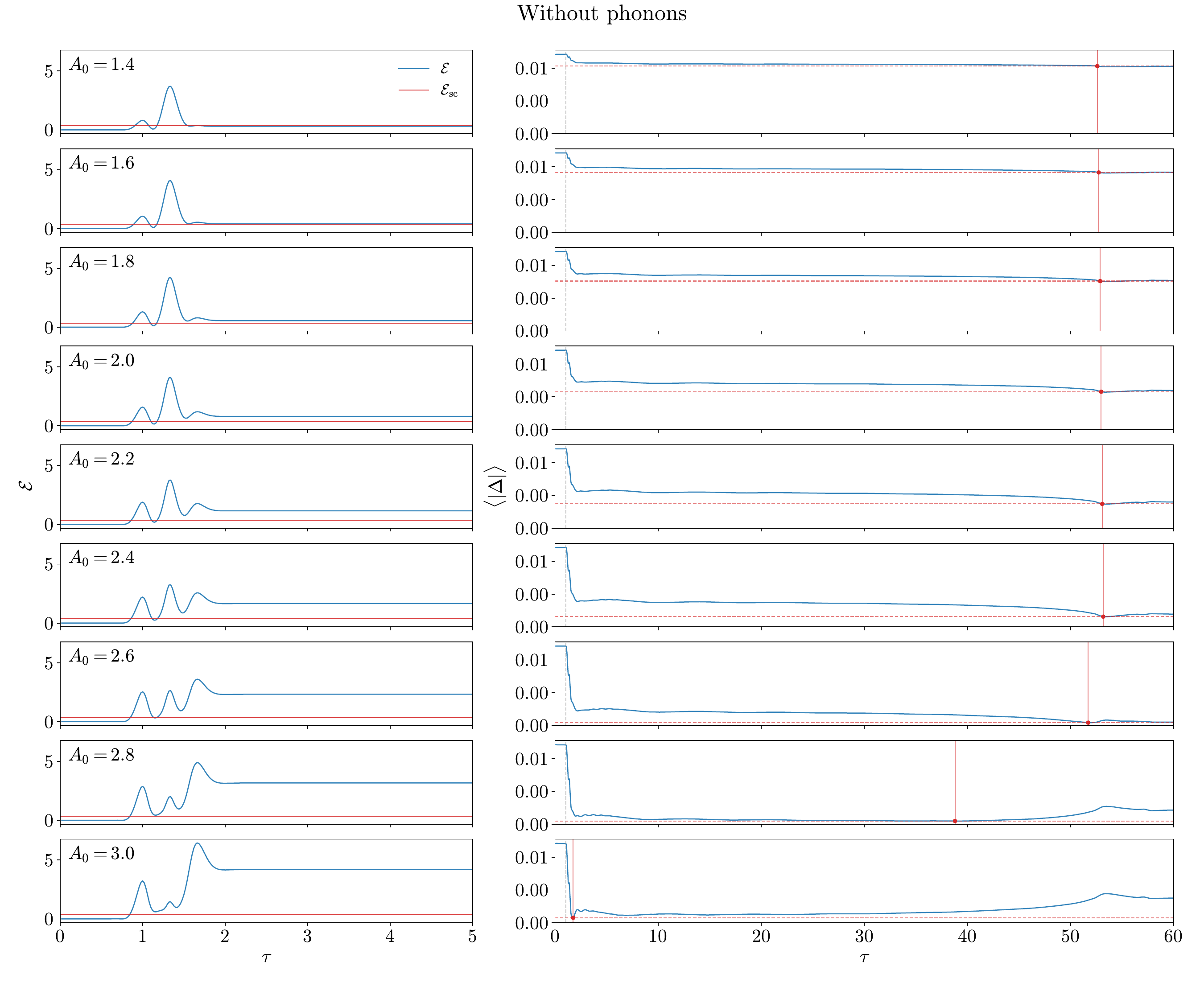}
	\caption{Time dynamics of the absorbed internal energy and the order parameter
	averaged over all sites with phonons. The leftmost plots shows the absorbed internal energy
	$\mathcal E$ in blue with gradually increasing laser amplitude $A_0$, with the horizontal red line marking the difference in
	energy between the superconducting and normal state before the laser pulse is
	applied. The rightmost plots shows the time evolution of the site-averaged order
	parameter $\expect{|\Delta|}$ in blue. The red dashed horizontal line marks
	the minimal value that $\expect{|\Delta|}$ attains during the time-evolution,
	and the solid red vertical line and dot shows the time $\tau_m$ where this minimum
	value is attained. Parameters used are $N_x = 100$, $\omega = 0.3$, $\gamma = 0.05$ and $\beta = 75$.
	}
	\label{fig:phonon_deltatime}
\end{figure*}

In Fig.~\ref{fig:tauvsdeltae_phonon} and Fig.~\ref{fig:rainbow_phonon}, the melting
time $\tau_m$ versus injected energy $\mathcal E_f$ and rainbow plot of the time
evolution of the site-averaged order parameter is shown. 
In Fig.~\ref{fig:phonon_deltatime}, the absorbed energy as a function of time is
shown next to the time evolution of the order parameter, marking the melting time
and its relation to the ratio of $\mathcal E_f$ and $\mathcal E_{\mr{sc}}$. Notably,
the absorbed energy for a given $A_0$ is much lower when coupling between electrons
and phonons are present than when it is absent. The behavior of the melting time
is qualitatively similar to the case without phonons, with the main difference
being the absolute numbers and the behavior below $\mathcal E_{\mr{sc}}$, where the
melting time changes less before the rapid decrease. The latter behavior is more so explained
by finite-size effects and quirks of the definition of melting time rather than a
physical effect.

\subsection{Periodic boundary conditions}

\begin{figure*}[p]
	\includegraphics[width=1\linewidth]{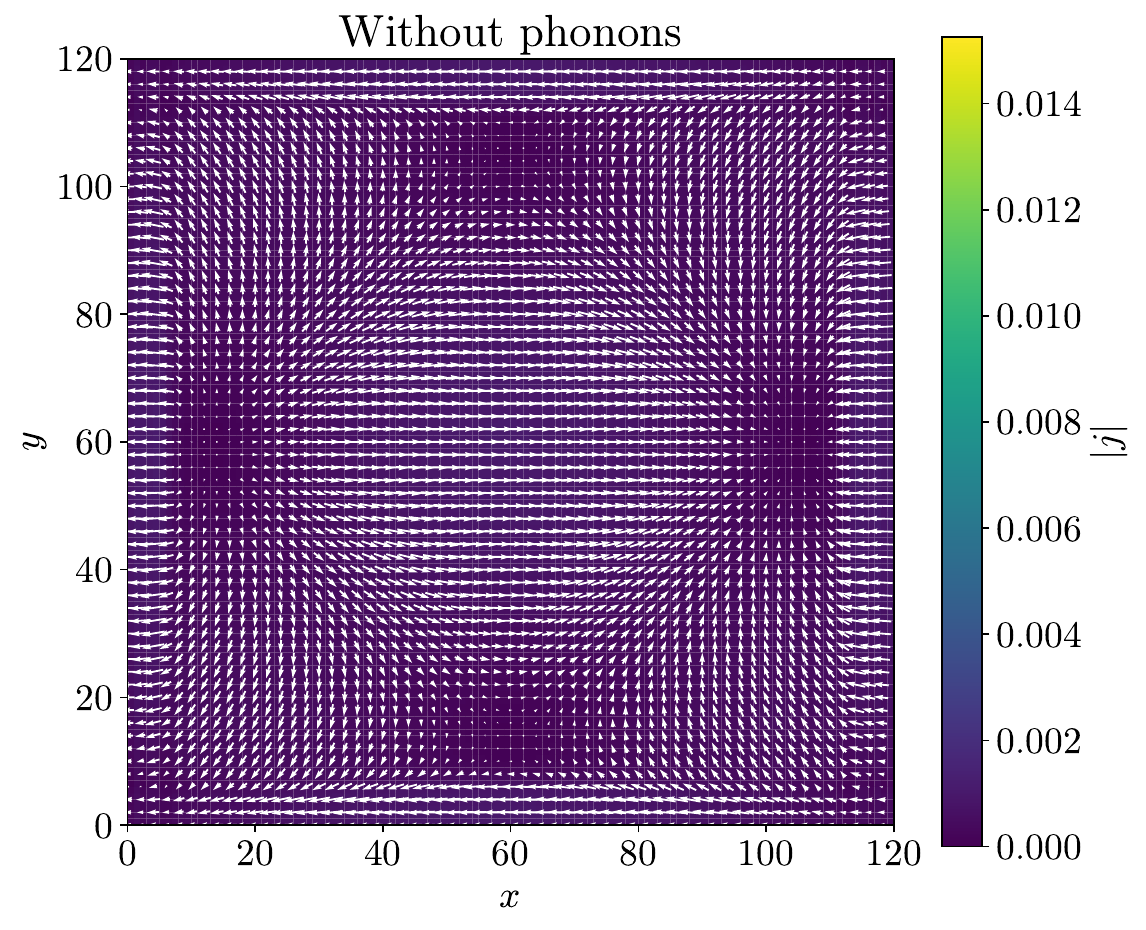}
	\caption{Currents on a lattice at time $\tau = 5.5$ with periodic boundary conditions and a Gaussian
	modulated vector potential. The two magnitude-wavefronts on either end are both moving
	towards the center of the plot, but the currents are in the $-x$-direction.
    Parameters used are $A_0 = 2.2$, $\beta = 10000$, $\mu = \omega = \gamma = 0$ and $\sigma_g = 60$.}
	\label{fig:pbc_currents}
\end{figure*}

The currents detailed in the main text depended on the hard boundary conditions
of the lattice. If we implement periodic boundary conditions without also considering
how the vector potential varies in space it is easy to see that the currents will vanish
since there are no special sites where electrons can accumulate. With spatial variation though,
for example a Gaussian modulation of the vector potential, currents are possible.
The Gaussian modulation is given by modifying the vector potential according to
\begin{equation}
    A(x, y, t) = A(t) e^{-\frac{(x - x_c)^2 + (y - y_c)^2}{2\sigma_g^2}},
\end{equation}
where $x_c$, $y_c$ corresponds to the center of the lattice and $\sigma_g$ decides the width of the modulation.
See \eg Fig.~\ref{fig:pbc_currents}, though note that the Gaussian modulation is
exaggerated compared to a realistic modulation in space to fit on the lattice and
for illustrative purposes.

\bibliography{references}